\documentclass[preprint,aps,tightenlines,showpacs,nofootinbib]{revtex4}
\usepackage{latexsym}
\usepackage[dvips]{graphicx}
\newcommand{\beq}{\begin{eqnarray}}
\newcommand{\eeq}{\end{eqnarray}}
\newcommand{\eq}{eqnarray}

\newcommand{\al}{{\alpha}}
\newcommand{\be}{{\beta}}
\newcommand{\ci}{\cite}

\newcommand{\de}{{\delta}}
\newcommand{\De}{\Delta}

\newcommand{\Om}{{\Omega}}

\newcommand{\no}{{\nonumber}}
\newcommand{\f}{\frac}
\newcommand{\ra}{\rightarrow}

\newcommand{\Sch}{Schwarzschild }

\begin{document}

\preprint{hep-th/0709.2307}

\title{The Generalized Uncertainty Principle in (A)dS Space and
the Modification of Hawking Temperature from the Minimal Length}

\author{Mu-In Park \footnote{E-mail address: muinpark@gmail.com}
}
\affiliation{ 
Research Institute of Physics and Chemistry, Chonbuk
National University, Chonju 561-756, Korea  \\
}
\begin{abstract}
Recently, the Heisenberg's uncertainty principle has been extended
to incorporate the existence of a large (cut-off) length scale in de
Sitter or anti-de Sitter space, and the Hawking temperatures of the
Schwarzshild-(anti) de Sitter black holes have been reproduced by
using the extended uncertainty principle. I generalize the extended
uncertainty to the case with an absolute minimum length and compute
its modification to the Hawking temperature. I obtain a {\it
general} trend that the generalized uncertainty principle due to the
absolute minimum length ``always'' increases the Hawking
temperature, implying ``faster'' decay, which is in conformity with
the result in the asymptotically flat space. I also revisit the {\it
black hole-string} phase transition, in the context of the
generalized uncertainty principle.
\end{abstract}

\pacs{04.70.Dy, 04.60.-m, 03.65.Ta}

\maketitle

\newpage

\section{Introduction}
The Heisenberg's uncertainty principle provides a basic limitation
of measuring the classical trajectories in the atomic or sub-atomic
scale. But here, there is no absolute minimum or maximum uncertainty
in the position and momentum themselves, though there is
``conditional'' minimum in them when one of them is fixed. So, in
this regards, there have been arguments that the Heisenberg's
uncertainty principle needs some modifications when the
gravitational interaction is considered in quantum mechanics since
there is an absolute minimum uncertainty in the position of any
gravitating quantum \ci{Wign:57,Amat:89}. And also, its several
interesting implications have been studied in the literatures.
Especially, it has been found that the generalized uncertainty
principle (GUP) increases the Hawking temperature, resulting in
``faster'' decay of Schwarzschild black holes in any dimension
\ci{Adle:01,Cava:04}.

However, the GUP does not have any limitation on the maximum
uncertainty in the position such as it can not be naively applied to
the case with the large (cut-off) length scales, like as in de
Sitter or anti-de Sitter space. Actually, the Hawking temperature of
black holes in (anti) de Sitter space can not be reproduced by the
Heisenberg's uncertainty principle or the GUP. Recently, an {\it
extended} uncertainty principle ( I will call this ``EUP'', simply)
has been introduced to incorporate the existence of the large length
scales and it is found that the Hawking temperatures of the
Schwarzshild-(anti) de Sitter black holes have been correctly
reproduced \ci{Bole:05}.

In this paper, I generalize the EUP to the case with an absolute
minimum uncertainty in the position as well and compute its
modification to the Hawking temperature. I obtain a {\it general}
trend that the generalized uncertainty principle due to the absolute
minimum length always increases the Hawking temperature, implying
faster decay, which is in conformity with the result of the
asymptotically flat space. I also revisit the black hole-string
phase transition, in the context of the generalized uncertainty
principle.

\section{The GUP and Hawking temperature in asymptotically flat space}
\label{2}

In this section, I review, with some new interpretations and
remarks, the GUP and the derivation of Hawking temperature from the
uncertainty principle in the asymptotically flat space
\ci{Adle:01,Cava:04}.

The GUP is given by
\begin{\eq}
\label{GUP}
 \De x_i \De p_j \ge \hbar \de_{ij}\left[ 1 + \al^2 l^2_P
\f{(\De p_j)^2}{\hbar ^2} \right],
\end{\eq}
where $x_i$ and $p_j ~(i,j=1,\cdots, d-1)$ are the spatial
coordinates and momenta, respectively; $l_P =(\hbar G)^{1/(d-2)}$ is
the Planck length and $\al$ is a dimensionless real constant of
order one \ci{Wign:57}. In the absence of the second term in the
right hand side, this reduces to the usual Heisenberg's uncertainty
principle without any ``absolute'' bound of $\De x_i$ nor $\De p_j$
themselves. But, in the presence of the second term, there exists an
absolute minimum in the position uncertainty
\begin{\eq}
\De x_i \geq 2 \al l_P
\end{\eq}
and the uncertainty in the momentum is given by
\begin{\eq}
\label{p:flat}
 \f{\hbar \De x_i}{2 \al^2 l_P^2} \left[1-
\sqrt{1-\f{4 \al^2 l_P^2}{(\De x_i)^2 }} \right] \leq \De p_i \leq
\f{\hbar \De x_i}{2 \al^2 l_P^2} \left[1+ \sqrt{1-\f{4 \al^2
l_P^2}{(\De x_i)^2 }} \right].
\end{\eq}
The left inequality in (\ref{p:flat}) provides some small
corrections to the Heisenberg's uncertainty principle for $\De x_i
\gg \al l_P$ (i.e., semi-classical regime),
\begin{\eq}
\label{p:flat(series)}
 \De p_i \geq \f{\hbar}{\De x_i} + \f{\hbar \al^2
l_P^2}{(\De x_i)^3} + {\cal O} \left(\f{\hbar \al^4 l_P^4}{(\De
x_i)^5 } \right).
\end{\eq}
On the other hand, the right inequality implies that $\De p_i$ can
not be arbitrarily large in order that the correction in (\ref{GUP})
makes sense. Of course, this upper bound can be higher with the
higher order terms in the right hand side of the GUP (\ref{GUP}),
but the absolute minimum in $\De x_i$ can be also lowered or even
disappeared, depending on the parameters \ci{Ko:06}. Another more
interesting interpretation would be that the upper bound corresponds
to the limit where the quantum gravity effects are very strong such
as a {\it black hole-string} phase transition can occur
\ci{Suss:93}. Actually, the inequality can be written also as
\begin{\eq}
\De p_i \leq \f{\hbar \De x_i}{ \al^2 l_P^2},
\end{\eq}
which can be directly derived also from the high momentum
uncertainty $\De p_j$ limit in (\ref{GUP}), and it is saturated by
the linear relation $\De p_i ={\hbar \De x_i}/{ \al^2 l_P^2}$, which
coincides with that of strings at the high energy limit, by
identifying the string scale $l_S \approx \al l_P$
\ci{Amat:89,Bole:05}.

Now, let me derive the Hawking temperature from the uncertainty
principle and general properties of black holes. To this end, let me
first consider a $d-$dimensional Schwarzshild black hole with a
metric given by
\begin{\eq}
\label{metric}
 ds^2=-N^2 dt^2 +N^{-2} dr^3 +r^2 d \Om^2_{d-2},
\end{\eq}
where
\begin{\eq}
N^2= 1-\f{16 \pi G M}{(d-2) \Om_{d-2} r^{d-3} }
\end{\eq}
and $\Om_{d-2}$ is the area of the unit sphere $S^{n-2}$
\ci{Myer:86}. By modeling a black hole as a black box with linear
size $r_+$, the uncertainty in the position of an emitted particle
by the Hawking effect is
\begin{\eq}
\label{Hawking effect}
 \De x_i \approx r_+
\end{\eq}
with the radius of the event horizon $r_+$. In the absence of the
GUP effect, the horizon radius is given by $r_+=[16 \pi GM/(d-2)
\Om_{d-2} ]^{1/(d-3)}$ from the metric (\ref{metric}). On the other
hand, in the presence of the GUP effect, the {\it precise} form of
the horizon radius $r_+=r_+ (M, \alpha l_P)$ is not known unless the
GUP corrected metric is known, which is beyond the scope of this
paper. However, I note that the relation (\ref{Hawking effect})
would be generally valid even with the GUP effect, with
understanding $r_+$ as the GUP corrected horizon already. Then, the
uncertainty in the energy of the emitted particle is ( by neglecting
the mass of the emitted particle )\footnote{There might exist high
energy modifications in the dispersion relation (\ref{Ep}),
generally \ci{Amel:06}. But, I will not consider this possibility in
this paper.}
\begin{\eq}
\label{Ep}
 \De E \approx \De p_i.
\end{\eq}
By assuming that $\De E$, which can be identified as the
characteristic temperature of the Hawking radiation, saturates the
left inequality \footnote{This assumption would correspond to the
Bekenstein bound of the entropy of an arbitrary bounded system $ S
(=\int T^{-1} dM) \leq S_{BH} (=\int T_{BH}^{-1} dM_{BH})$ whose
upper bound is saturated by that of black holes, $S_{BH}$, for a
given mass $M=M_{BH}$\ci{Beke:81}.}, one can obtain the Hawking
temperature
\begin{\eq}
\label{GUP:exact(flat)}
 T_{GUP}=``\left( \f{d-3}{4 \pi} \right) "
\f{\hbar r_+}{2 \al^2 l_P^2}\left[1- \sqrt{1-\f{4 \al^2 l_P^2}{r_+^2
}} \right].
\end{\eq}
Here, the ``calibration'' factor `$(d-3)/4 \pi$' has been introduced
in order to have agreements with the usual Hawking temperature of
the \Sch black hole in the leading term, for a large black hole ,
i.e., $r_+ \gg \alpha l_P $ \ci{Hawk:75,Myer:86}:
\begin{\eq}
\label{GUP:series(flat)}
 T_{GUP}=\left( \f{d-3}{4 \pi} \right)\left[
\f{\hbar}{ r_+} + \f{\hbar \al^2 l_P^2}{r_+^3} + {\cal O}
\left(\f{\hbar \al^4 l_P^4}{r_+^5 }\right) \right].
\end{\eq}
Before finishing this section, I remark first that the formula
(\ref{Hawking effect}), as a result (\ref{GUP:exact(flat)}), is
still valid even for the small black holes up to the absolute
minimum, which is order of Planck length $l_P$, though the series
formula (\ref{GUP:series(flat)}) is valid only for a large $r_+$.
The black hole evaporation stops at $r_+=2 \al l_P$, where the curve
ends, and this would correspond to a ``melting'' of the black hole
which is followed by the string phase, according to the new
interpretation \ci{Suss:93}. Second, the effect of the GUP with an
absolute minimum length increases the Hawking temperature always and
this implies that it decays faster than the usual \Sch black hole
without the GUP (Fig.1).

\begin{figure}
\includegraphics[width=7cm,keepaspectratio]{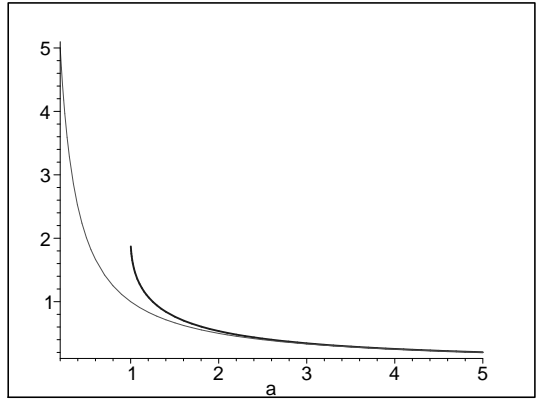}
\caption{Hawking temperature (divided by `$(d-3)/4 \pi$') vs. the
horizon radius $r_+$ (denoted by `$a$' in the plot) in the
asymptotically flat space. In the absence of the GUP, there is no
absolute minimum radius for the black hole evaporation (thin line).
With the GUP, the Hawking temperature becomes hotter, implying
faster decay, and also there is a minimum radius $r_+=2 \al l_P$
where the curve ends, implying that the black hole evaporation stops
(thick line). Here, I have plotted the cases with
$\hbar=l_P=1,~\al=0.5$ and the GUP curve stops at $r_+=1$. }
\end{figure}

\section{The EUP and Hawking temperature in (A)dS space}

The GUP can not be naively applied to the space with the large
length scales like as in (A)dS space \footnote{This has been noted
earlier by Konishi et al. also \ci{Amat:89}. See also Ref.
\cite{Myung:07} for another related work. }. In this section, I
consider an extension of the uncertainty principle in order to
incorporate the large-length scales and derivation of Hawking
temperature from the uncertainty principle.

The extended uncertainty principle (EUP) is given by \footnote{The
parameter $\beta$ in Ref. \ci{Bole:05} is related to here's by
$\beta_{There}=(l_P/l) \beta_{Here}$.}
\begin{\eq}
\label{EUP}
 \De x_i \De p_j \ge \hbar \de_{ij}\left[ 1 + \be^2
\f{(\De x_i)^2}{l^2} \right],
\end{\eq}
where $l$ is the characteristic, large length scale and $\be$ is a
dimensionless real constant of order one \ci{Bole:05}\footnote{This
has been considered earlier by Kempf et. al. also \ci{Kemp:95}, but
its physical consequences have not been studied.}. (For some {\it
gedanken} experiments' derivation, even without considering black
holes, see also Ref. \ci{Bamb:07}.) Then, it is easy to see that
there is an absolute minimum in the momentum uncertainty
\begin{\eq}
\label{EUP:2}
 \De p_i \ge \f{\hbar}{\De x_i }+\f{ \hbar \be^2
\De x_i}{l^2} \ge \f{2 \hbar \be}{l}.
\end{\eq}
Here, I note that the first inequality is an ``exact'' relation
drawn from (\ref{EUP}), without considering any limit as in
(\ref{p:flat(series)}).

Now, using the approach in Sec. II, it is straightforward to see
that the Hawking temperature of the Schwarzshild-AdS black holes
from the EUP (\ref{EUP:2}) \footnote{For an alternative derivation
from the laws of classical physics and Heisenberg's uncertainty
principle, see Ref. \ci{Scar:06}. But, there is no room for the GUP
in that derivation.}. To this end, let me first consider a
$d$-dimensional Schwarzshild-AdS black hole with the metric function
\begin{\eq}
N^2= 1+ \f{r^2}{l^2_{AdS}}-\f{16 \pi G M}{(d-2) \Om_{d-2} r^{d-3} }
\end{\eq}
in the metric (\ref{metric}) and a cosmological constant
$\Lambda=-(d-1)(d-2)/2 l^2_{AdS}$ \ci{Hawk:83}. Then, with the same
identifications (\ref{Hawking effect}) and (\ref{Ep}) for the
Hawking-emitted particles, which do not depends on the large scale
behaviors but only on the local structure near the horizon, one can
obtain the Hawking temperature $T_{EUP} \approx \De p_i$,
\begin{\eq}
\label{EUP:exact}
 T_{EUP(AdS)}=\left( \f{d-3}{4 \pi} \right)
\hbar \left[ \f{1}{r_+}+ \left(\f{d-1}{d-3}\right)
\f{r_+}{l^2_{AdS}} \right]
\end{\eq}
with the same calibration factor `$(d-3)/4 \pi$' as in the
asymptotically flat case, implying its universality, and
$\be=\sqrt{(d-1)/(d-3)}, ~l=l_{AdS}$; $r_+$ is the radius of the
event horizon which solves $N^2(r)=0$. Here, the existence of the
absolute minimum in $\De p_i$ and so in $T_{EUP(AdS)}$ is a general
consequence of the EUP of (\ref{EUP}) (see Fig. 2 (thin line)).

So far, I have shown that the EUP in (\ref{EUP}) applies to the AdS
space. Now, the EUP for the dS space can be easily constructed by
considering $l^2 \ra -l^2$ in (\ref{EUP}):
\begin{\eq}
\label{EUP:dS}
 \De x_i \De p_j \ge \hbar \de_{ij}\left[ 1 - \be^2
\f{(\De x_i)^2}{l^2} \right].
\end{\eq}
Then, in contrast to (\ref{EUP}), there is an absolute maximum in
$\De x_i$ as
\begin{\eq}
\label{x:upper}
 \De x_i  \le \f{l}{\beta}
\end{\eq}
in order that $\De p_i$ is not negative\footnote{This is compared
with Refs. \ci{Bamb:07,Noza:06}, where the EUP (\ref{EUP}) is
considered for the particle or cosmological horizon, by giving the
absolute minimum in $\Delta p_i$. },
\begin{\eq}
 \De p_i \ge \f{\hbar}{\De x_i }-\f{ \hbar \be^2
\De x_i}{l^2}  \ge 0.
\end{\eq}
Note that the absolute maximum in $\De x_i$ does not have $\hbar$
such as this is a purely classical result.

The Hawking temperature of the Schwarzshild-dS black hole with a
cosmological constant $\Lambda=+(d-1)(d-2)/2 l^2_{dS}$ \ci{Gibb:77}
is similarly computed as, by considering $l^2_{AdS} \ra -l^2_{dS}$
in (\ref{EUP:exact}),
\begin{\eq}
\label{EUP:exact(dS)}
 T_{EUP(dS)}=\left( \f{d-3}{4 \pi} \right)
\hbar \left[ \f{1}{r_+}- \left(\f{d-1}{d-3}\right) \f{r_+}{l^2_{dS}}
\right].
\end{\eq}
Here, the maximum bound reads ( $l=l_{dS},\be=\sqrt{(d-1)/(d-3)}$ )
\begin{\eq}
\label{N:horizon}
 r_+ \leq \sqrt{(d-1)/(d-3)} l_{dS},
\end{\eq}
which is the Nariai bound where the black hole horizon and the
cosmological horizon meet \ci{Nari:51}. So, the condition
(\ref{x:upper}) reflects the fact that the uncertainty in the
position can not exceed the cosmological horizon, which is the size
of the casually connected world in a dS space (see Fig.3 (thin
line)).

\section{The generalized EUP (GEUP)}

In the EUP (\ref{EUP}), there is an absolute minimum in the
uncertainty of the momentum. In this section, I generalize the EUP
to have a minimum length scale as well, by combining the GUP and the
EUP, and study the effect of the minimum length to the Hawking
temperature from the EUP, i.e., the Hawking temperature of
Schwarzshild-(A)dS black holes.

The generalized EUP ($G$EUP) is given by
\begin{\eq}
\label{GEUP}
 \De x_i \De p_j \ge \hbar \de_{ij}\left[ 1 +\al^2 l^2_P
\f{(\De p_j)^2}{\hbar ^2} + \be^2 \f{(\De x_i)^2}{l^2} \right],
\end{\eq}
where I have considered the case of the AdS space, first. Then, by
inverting (\ref{GEUP}), one has the inequalities,
\begin{\eq}
\label{p:GEUP}
&& \De p_i^{(-)} \leq \De p_i \leq \De p_i^{(+)}, \no \\
&& \De p_i^{(\pm)}=\f{\hbar \De x_i}{2 \al^2 l_P^2} \left[1 \pm
\sqrt{1-\f{4 \al^2 l_P^2}{(\De x_i)^2 } \left[ 1+\be^2 \f{(\De
x_i)^2}{l^2} \right]} \right]
\end{\eq}
and
\begin{\eq}
\label{x:GEUP}
&& \De x_i^{(-)} \leq \De x_i \leq \De x_i^{(+)}, \no \\
&& \De x_i^{(\pm)}=\f{l^2 \De p_i}{2 \hbar \be^2} \left[1 \pm
\sqrt{1-\f{4 \be^2 \hbar^2}{ l^2 (\De p_i)^2 } \left[ 1+\f{ \al^2
l_P^2 (\De p_i)^2}{\hbar^2} \right]} \right].
\end{\eq}
Here, one finds that there are, now, both the absolute minimum in
$\De x_i$ and $\De p_i$
\begin{\eq}
\label{x:lower:AdS}
&&(\De x_i)^2 \geq \f{4 \al^2 l_P^2}{1-4 \al^2 l_P^2 \be^2/l^2}~, \\
&&(\De p_i)^2 \geq \f{4 \hbar^2 \be^2/ l^2}{1-4 \al^2 l_P^2
\be^2/l^2}
\end{\eq}
from the {\it reality} of $\De p^{(\pm)}_i$ and $\De x^{(\pm)}_i $,
respectively, with the condition
\begin{\eq}
\be^2 < \f{l^2}{4 \al^2 l_P^2}.
\end{\eq}
The left inequality in (\ref{p:GEUP}), as in (\ref{p:flat}) of the
GUP, provides some small corrections to the Heisenberg's uncertainty
principle, due to the minimum length and momentum, for $\al l_P \ll
\De x_i \ll l/\be$,
\begin{\eq}
 \De p_i \ge \left( 1+ \f{2  \al^2 l_P^2 \be^2
}{l^2} \right) \f{\hbar}{\De x_i }+\f{ \hbar \be^2 \De
x_i}{l^2}+\f{\hbar \al^2 l_P^2}{(\De x_i)^3} + {\cal O}
\left(\f{\hbar \al^4 l_P^4}{(\De x_i)^5}, \f{\hbar \al^2 l_P^2 \be^4
\De x_i}{l^4} \right) .
\end{\eq}
By repeating the same arguments as in the GUP and the EUP cases
(with understanding that $r_+$ as the GUP corrected horizon), one
can obtain the Hawking temperature $T_{GEUP} \approx \De p^{(-)}_i$
\begin{\eq}
\label{GEUP:exact}
 T_{GEUP(AdS)}=\left( \f{d-3}{4 \pi} \right)
\f{\hbar r_+}{2 \al^2 l_P^2} \left[1- \sqrt{1-\f{4 \al^2
l_P^2}{r_+^2 } \left[1+ \left(\f{d-1}{d-3}\right)
\f{r_+^2}{l^2_{AdS}} \right]} \right]
\end{\eq}
with the usual calibration factor `$(d-3)/4\pi$' and
$\be=\sqrt{(d-1)/(d-3)}, ~l=l_{AdS}$ such as this agrees with the
EUP result (\ref{EUP:exact}) for a semiclassical black hole with
$\al l_P \ll r_+ \ll  \sqrt{(d-3)/(d-1)} l_{AdS}$,
\begin{\eq}
\label{GEUP:AdS}
 T_{GEUP(AdS)} \approx \left( \f{d-3}{4 \pi} \right)
\hbar \left[ \left\{1+ \left(\f{d-1}{d-3} \right) \f{2  \al^2
l_P^2}{l_{AdS}^2} \right\}\f{1}{r_+}+ \left(\f{d-1}{d-3} \right)
\f{r_+}{l^2_{AdS}} + \f{\al^2 l_P^2}{r_+^3} \right].
\end{\eq}
Here, the third term is purely the GUP correction and the second
term in the first bracket $\{  ~\}$ is the {\it G}EUP effect, and
these correction terms are all positive. This shows that the Hawking
temperature of the AdS black hole is increased also by the minimum
uncertainty in the position, with the GUP.

\begin{figure}
\includegraphics[width=7cm,keepaspectratio]{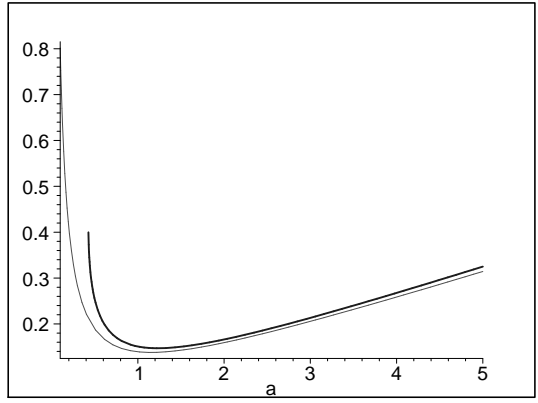}
\caption{Hawking temperature vs. the horizon radius $r_+$in the AdS
space. The EUP, but without the GUP, produces correctly the usual
Hawking temperature of the Schwarzshild-AdS black holes (thin line).
The existence of the absolute minimum in the temperature is a
general consequence of the EUP. But, as in the case of the flat
space, there is no absolute minimum radius in the absence of the
GUP. With the GUP, the Hawking temperature becomes hotter also,
implying faster decay, and there is a minimum radius $r_+=2 \al l_P/
[1-4 \al^2 l_P^2 (d-1)/(d-3)l_{AdS}^2 ]^{1/2}$ where the curve ends,
implying that the black hole evaporation stops (thick line). Here, I
have plotted the cases with $\hbar=l_P=1,~\al=0.2,~l_{AdS}=2,~d=4$.}
\end{figure}

The analysis for the dS case is also straightforward. From the GEUP
with $l^2 \ra -l^2$, one has
\begin{\eq}
\label{p:GEUP(dS)}
&& \De p_i^{(-)} \leq \De p_i \leq \De p_i^{(+)}, \no \\
&& \De p_i^{(\pm)}=\f{\hbar \De x_i}{2 \al^2 l_P^2} \left[1 \pm
\sqrt{1-\f{4 \al^2 l_P^2}{(\De x_i)^2 } \left[ 1-\be^2 \f{(\De
x_i)^2}{l^2} \right]} \right]
\end{\eq}
with the minimum uncertainty in $\De x_i$ (but none in $\De p_i$)
\begin{\eq}
\label{x:lower:GEUP(dS)}
 (\De x_i)^2 \geq \f{4 \al^2 l_P^2}{1+4
\al^2 l_P^2 \be^2/l^2}.
\end{\eq}
Moreover, in order that $\De p^{(-)}_i$ is not negative one obtains
the same condition as (\ref{x:upper}) which is unchanged by the GUP
effect (i.e., no $\al$ dependence), in contrast to the lower bound
in (\ref{x:lower:GEUP(dS)}). Then, one finds the Hawking temperature
\begin{\eq}
\label{GEUP:exact(dS)}
 T_{GEUP(dS)}=\left( \f{d-3}{4 \pi} \right)
\f{\hbar r_+}{2 \al^2 l_P^2} \left[1- \sqrt{1-\f{4 \al^2
l_P^2}{r_+^2 } \left[1- \left(\f{d-1}{d-3}\right)
\f{r_+^2}{l^2_{dS}} \right]} \right]~,
\end{\eq}
which gives
\begin{\eq} \label{GEUP:dS}
 T_{GEUP(dS)} \approx \left( \f{d-3}{4 \pi} \right)
\hbar \left[ \left\{1- \left(\f{d-1}{d-3} \right) \f{2  \al^2
l_P^2}{l_{dS}^2} \right\}\f{1}{r_+}- \left(\f{d-1}{d-3} \right)
\f{r_+}{l^2_{dS}} + \f{\al^2 l_P^2}{r_+^3} \right]
\end{\eq}
for semiclassical dS black holes with $\al l_P \ll r_+ \ll
\sqrt{(d-3)/(d-1)} l_{dS}$. Here, note that the maximum bound of the
black hole horizon (\ref{N:horizon}) is not changed by the existence
of the minimal length but the temperature is always increasing: The
second term in the first bracket $\{  ~\}$ gives a negative
correction but this is dominated by the third term, which is always
positive.

\begin{figure}
\includegraphics[width=7cm,keepaspectratio]{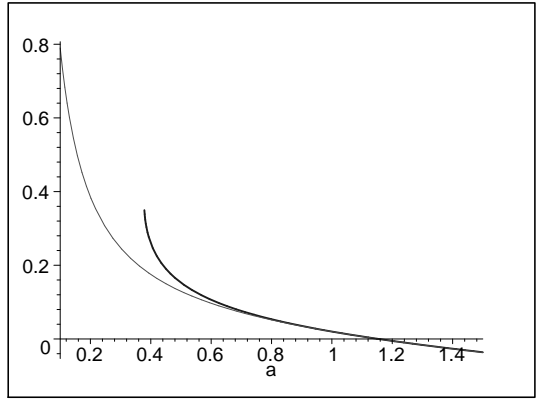}
\caption{Hawking temperature vs. the horizon radius $r_+$in the dS
space. The EUP of (\ref{EUP:dS}) produces correctly the usual
Hawking temperature of the Schwarzshild-dS black holes (thin line),
which vanishes at the Nariai bound $r_+ \leq \sqrt{(d-1)/(d-3)}
l_{dS}$; this defines the absolute maximum of the black hole
horizon, but there is no absolute minimum radius. With the GUP, the
Hawking temperature becomes hotter also, implying faster decay, and
there is a minimum radius $r_+=2 \al l_P/ [1+4 \al^2 l_P^2
(d-1)/(d-3)l_{dS}^2]^{1/2}$ where the curve ends, implying that the
black hole evaporation stops (thick line). Here, I have plotted the
cases with $\hbar=l_P=1,~\al=0.2,~l_{dS}=2,~d=4$.}
\end{figure}

Now, one finds a quite general trend that the GUP due to a minimal
length increases {\it always} the Hawking temperature (Fig.2, 3),
regardless of being asymptotically flat or (A)dS space. This can be
traced back to the {\it universal} appearance of the term ``$+\al^2
l_P /r_+^3$'' in the temperature formula, which makes the decay to
be faster. This seems to be also true in other forms of the
deformation of the uncertainty principle \ci{Park:tba}.

Finally, two remarks are in order. First, one might consider the
first law of thermodynamics to compute the GUP corrected black hole
entropy from the {\it same} ADM mass formula as that of the case
without the GUP \ci{Adle:01}. But, it still unclear how to fix {\it
uniquely} the GUP corrected mass formula from the GUP corrected
Hawking temperature, without knowing the precise form the GUP
corrected gravity and its black hole solutions.

Second, I note that, in the $d=3$ limit of the AdS black holes
(i.e., the BTZ black hole limit), one has the Hawking temperature
\begin{\eq}
\label{GEUP:AdS3}
 T^{(d=3)}_{GEUP(AdS)} \approx \f{\hbar}{4 \pi}
\left[  \f{4  \al^2 l_P^2}{l_{AdS}^2} \f{1}{r_+}+ \f{2
r_+}{l^2_{AdS}} \right]
\end{\eq}
from the series formula (\ref{GEUP:AdS}), though it needs a scale
tuning $l_{AdS} \ra \infty, d\ra 3$, with `$\sqrt{d-3} l_{AdS} $= a
fixed large number'. This shows also an increase of the temperature,
implying faster decay from the GUP effect, compared to that of the
usual BTZ black hole, $T_{BTZ}=\hbar r_+ /(2 \pi l^2_{AdS})$. But,
remarkably, there is a minimum temperature at $r_+=\sqrt{2} \al l_P$
and growing temperature for smaller black holes, in contrast to the
monotonically decreasing temperature as $r_+$ becomes smaller in the
BTZ black hole without the GUP. If this were true, the Hawking-Page
transition \ci{Hawk:83} would occur even in three-dimensional AdS
space, due to the GUP effect. But, this does not seem to occur from
(\ref{x:lower:AdS}), which implies $r_+ \geq 2 \al l_P $ for
consistency of the exact formula (\ref{p:GEUP}), such as the
evaporation stops before reaching the absolute minimum of the
temperature at $r_+=\sqrt{2} \al l_P$. This needs more rigorous
analysis which can be well-defined in the three dimension, from the
start \ci{Park:tba}.


\section*{Acknowledgments}

I would like to thank Drs. Yong-Wan Kim, Sunggeun Lee, Profs. Yun
Soo Myung, Young-Jai Park, and Chaiho Rim for helpful discussions
about the GUP. I would like to also thank the hospitality of CQUeST
for providing the facility during this work.
This work was supported by the Korea Research Foundation Grant
funded by Korea Government(MOEHRD) (KRF-2007-359-C00011)

\newcommand{\J}[4]{#1 {\bf #2} #3 (#4)}
\newcommand{\andJ}[3]{{\bf #1} (#2) #3}
\newcommand{\AP}{Ann. Phys. (N.Y.)}
\newcommand{\MPL}{Mod. Phys. Lett.}
\newcommand{\NP}{Nucl. Phys.}
\newcommand{\PL}{Phys. Lett.}
\newcommand{\PR}{Phys. Rev. D}
\newcommand{\PRL}{Phys. Rev. Lett.}
\newcommand{\PTP}{Prog. Theor. Phys.}
\newcommand{\hep}[1]{ hep-th/{#1}}
\newcommand{\hepp}[1]{ hep-ph/{#1}}
\newcommand{\hepg}[1]{ gr-qc/{#1}}
\newcommand{\bi}{ \bibitem}

\end{document}